\documentstyle[11pt,epsfig]{article}
\begin{document}
\title{Nonpartonic components in the nucleon structure functions
at small $Q^2$ in a broad range of $x$}

   \author{A. Szczurek$^{1}$ and V. Uleshchenko$^{1,2}$  \\
   {\it $^{1}$ Institute of Nuclear Physics, PL-31-342 Cracow, Poland  } \\
   {\it $^{2}$ Institute for Nuclear Research, 252028 Kiev, Ukraine  } \\   }
\maketitle
\begin{abstract}
We construct a simple two-phase model of the nucleon structure
functions valid for both small and large $Q^2$ and in the broad
range of Bjorken $x$.
The model incorporates hadron dominance at small $x$
and $Q^2$ and parton model at large $Q^2$.
The VDM contribution is modified for small fluctuation times of
the hadronic state of the photon.
With two free parameters we describe SLAC,
CERN NMC, Fermilab E665 and CERN BCDMS data for both proton and
deuteron structure functions.
Our model explains some
phenomenological higher-twist effects extracted from earlier analyses.
A good description of
the NMC $F_2^p(x) - F_2^n(x)$ data is obtained in contrast to other
models in the literature.
We predict faster vanishing of the partonic component at low $Q^2$ than
previously expected and strong $Q^2$ dependence of the Gottfried Sum
Rule below $Q^2 \approx$ 4 GeV$^2$.
\end{abstract}

\section{Introduction}

\mbox{}\indent
The standard deep inelastic scattering picture applies when
the four-momentum transfer squared from the lepton line to
the hadron line ($Q^2$) is large. When virtual photon wave length
increases and reaches the size of the nucleon one may expect a transition
to another regime where the standard partonic model is no longer valid.
In this region a kinematical constraint \cite{BK_lowq2} guarantees
the vanishing of the $F_2(x,Q^2)$ structure function. This requirement
is not embodied in the perturbative parton distributions.
A phenomenological fit based on parton screening was proposed
in \cite{MRS_lowQ2} to satisfy this condition by introducing
an extra form factor.

The recent low-$Q^2$ data from HERA \cite{H1_lowQ2,ZEUS_lowQ2}
have triggered many phenomenological analyses. Especially interesting
is the unexplored transition region. At present there is no consensus
on the details of the transition mechanism. Here we concentrate
on the region of somewhat larger $x$ rather than the new HERA data.
We shall demonstrate that also at larger $x$ a similar transition due to
vanishing partonic components at small $Q^2$
takes place, although it is not directly seen from experimental data.

It is common wisdom that the vector dominance model applies
at low $Q^2$ while the parton model
describes the region of large $Q^2$, leading at lowest order to
Bjorken scaling, and to logarithmic scaling violation in higher orders
of QCD. A proposal was made in Ref.\cite{BK_model} to unify both
the limits in a consistent dispersion method approach.
In the traditional formulation of the VDM one is limited to large
lifetimes
of hadronic fluctuations of the virtual photon, i.e. small Bjorken
$x < 0.1$ for the existing data.
It is a purpose of this paper to generalize the model
to a full range of $Q^2$ and $x$ by introducing extra phenomenological
form factors to be adjusted to the experimental data.

Some authors believe that it is old-fashioned to talk about
VDM contribution in the QCD era.
However, VDM effects appear naturally in the time-like region in
the production of vector mesons in $e^+ e^-$ collisions.
These effects cannot be described in terms of perturbative
QCD, as in the production of resonances many complicated
nonperturbative effects take place. The physics must be similar in
the space-like region.
We shall demonstrate that it is essential to include this contribution
explicitly in order to describe the structure functions at low Q$^2$.

In the next section we outline our model and discuss how to
choose its basic parameters in a model independent way.
In section 3 we discuss a fit of the remaining parameters
to the fixed target data and present results of the fitting procedure
for the proton and deuteron structure functions.
In addition we compare the result of our model for
$F_2^p - F_2^n$ and some subtle isovector higher-twist effects
with another low-$Q^2$ model. Finally we discuss some interesting
predictions which could be verified in the future.

\section{The model}

\mbox{}\indent
As in Ref.\cite{BK_model} the total nucleon structure function is
represented as a sum
of the standard vector dominance part, important at small $Q^2$ and/or
small Bjorken $x$, and
partonic ({\it part}) piece which dominates over the vector dominance
(VDM) part at large $Q^2$:
\begin{equation}
F_2^N(x,Q^2) = F_2^{N,VDM}(x,Q^2) + F_2^{N,part}(x,Q^2) \; .
\label{F2_decomposition}
\end{equation}
The standard range of applicability of vector dominance contribution
is limited to large invariant masses of the
hadronic system ($W$), i.e. small values of $x$.
In the target (nucleon) reference frame the time of life of the hadronic
fluctuation is given according to the uncertainty principle
as $\tau \sim 1/\Delta E$ with
\begin{equation}
\Delta E = \sqrt{M_V^2 + |{\bf q}|^2 } - \sqrt{q^2 + |{\bf q}|^2 } \; ,
\end{equation}
where $M_V$ is the mass of the hadronic fluctuation (vector meson mass).
In terms of the photon virtuality and Bjorken $x$ this can be expressed
as
\begin{equation}
\Delta E = \frac{M_V^2 + Q^2}{Q^2} \cdot M_N x  \; .
\end{equation}
As the energy transfer $\nu \rightarrow \infty$ the time of life
of the hadronic
fluctuaction becomes $\tau \sim \frac{Q^2}{M_V^2 m_N x}$.
It is natural to expect small VDM contribution when the time of
life of the hadronic fluctuation is small.
We shall model this fact by introducing a form factor
$\Omega(\tau) = \Omega(x,Q^2)$. Then the modified
vector dominance contribution can be written as:
\begin{equation}
F_2^{N,VDM}(x,Q^2) =
\frac{Q^2}{\pi} \sum_V
\frac{M_V^4 \cdot \sigma_{VN}^{tot}(s^{1/2})}
{\gamma_V^2 (Q^2 + M_V^2)^2}
\cdot \Omega_V(x,Q^2) \; .
\label{F2_VDM}
\end{equation}
In the present paper we take $\gamma$'s calculated from the leptonic
decays of vector mesons which include finite width corrections \cite{IKL}
$\gamma_{\rho}^2/4\pi$ = 2.54,
$\gamma_{\omega}^2/4\pi$ = 20.5,
$\gamma_{\phi}^2/4\pi$ = 11.7.
\footnote{Please note different normalization of $\gamma$'s
in comparison to \cite{BK_model}.}

In general one can try different functional forms for $\Omega$.
In the present analysis we shall use only exponential and Gaussian
form factors
\begin{eqnarray}
\Omega(x,Q^2) = \exp(-\Delta E / \lambda_E) \;, \nonumber \\
\Omega(x,Q^2) = \exp(-(\Delta E / \lambda_G)^2) \; .
\label{time_formfactor}
\end{eqnarray}
As in Ref.\cite{BK_model} we take the partonic contribution as
\begin{equation}
F_2^{N,part}(x,Q^2) =
\frac{Q^2}{Q^2 + Q_0^2} \cdot F_2^{asymp}(\bar x,\bar Q^2) \; .
\label{F2_part}
\end{equation}
where $\bar x = \frac{Q^2 + Q_2^2}{W^2 - m_N^2 + Q^2 + Q_2^2}$
and $\bar Q^2 = Q^2 + Q_1^2$.
The $F_2^{asymp}(x,Q^2)$ above denotes the standard partonic
structure function which in the leading order can be expressed
in terms of the quark distributions:
\begin{center}
$F_2^{asymp}(x,Q^2) = x \cdot \sum_f \; e_f^2 \cdot
 \left[ q_f(x,Q^2) + \bar q_f(x,Q^2) \right] $.
\end{center}
The extra factor in front of Eq.(\ref{F2_part}) assures a correct
kinematic beheviour in the limit $Q^2 \rightarrow 0$.
In general $Q_0^2$, $Q_1^2$ and $Q_2^2$ can be slightly different.
In the following section we shall consider different options.

At large Bjorken $x$ one has to include also the so-called target
mass corrections. Their origin is mainly kinematic \cite{GP76}.
In our approximate treatment we substitute the Bjorken variable $x$
in the partonic distributions by the Nachtmann variable $\xi$
\cite{Nachtmann}
given by:
\begin{equation}
\xi = \frac{2 x}
{1 + \sqrt{(1 + \frac{4 M_N^2 x^2}{Q^2})} } \; ,
\label{Nachtmann_variable}
\end{equation}
which is the dominant modification.

In principle $F_2^{asymp}(x,Q^2)$ could be obtained in any
realistic model of the nucleon combined with QCD evolution.
We leave the rather difficult problem of modeling the partonic
distributions for future studies.
We expect that at not too small $x >$ 0.01, the region of the interest
of the present paper, the leading order Gl\"uck-Reya-Vogt (GRV)
parametrization of
$F_2^{p,asymp}(x,Q^2)$ and $F_2^{n,asymp}(x,Q^2)$ should be
adequate. Furthermore in our opinion the parametrization \cite{GRV95}
with the valence-like input for the sea quark distributions and
$\bar d$ - $\bar u$ asymmetry built in incorporates in a
phenomenological way nonperturbative effects caused by the meson cloud
in the nucleon \cite{meson_cloud}.

The total cross section for (vector meson) -- (nucleon) collision
is not well known.
Above meson-nucleon resonances, one may expect
the following approximation to hold:
\begin{eqnarray}
\sigma_{\rho N}^{tot} &=& \sigma_{\omega N}^{tot}
= \frac{1}{2} \left[
\sigma_{\pi^{+} p}^{tot} + \sigma_{\pi^{-} p}^{tot} \right]
\; , \nonumber \\
\sigma_{\phi N}^{tot} &=&
\sigma_{K^+ p}^{tot} + \sigma_{K^- p}^{tot} -
\frac{1}{2} \left[
\sigma_{\pi^{+} p}^{tot} + \sigma_{\pi^{-} p}^{tot} \right] \; .
\end{eqnarray}
Using a simple Regge-inspired parametrizations
by Donnachie-Landshoff \cite{DL92}
of the total $\pi N$ and $K N$ cross sections we get simple
and economic parametrizations
for energies above nucleon resonances
\begin{eqnarray}
\sigma_{\rho N}^{tot} &=& \sigma_{\omega N}^{tot}
= 13.63 \cdot s^{0.0808} + 31.79 \cdot s^{-0.4525} \; , \nonumber \\
\sigma_{\phi N}^{tot} &=&
10.01 \cdot s^{0.0808} + 2.72 \cdot s^{-0.4525} \; ,
\end{eqnarray}
where the resulting cross sections are in mb.

We expect that our model should be valid in a broad range of
$x$ and $Q^2$ except for very small $x < 0.001$, where genuine effects
of BFKL pomeron physics could show up, and except for very large $x$,
where the energy ($s^{1/2}$ in Eq.(\ref{F2_VDM})) is small and
the behaviour of the total $VN$ cross section is
essentially unknown. Because our main interest is in the transition
region, the large $Q^2$ data were not taken into account
in the fit. There the partonic contribution is by far dominant and
the GRV parametrization \cite{GRV95} is known to provide a reliable
description of the data.


\section{Comparison to experimental data}

\mbox{}\indent
Most of the previous parametrizations in the literature
centered on the proton structure function.
In the present analysis we are equally interested in both proton
and neutron structure functions. Achieving this goal requires
special selection of the experimental data with similar statistics
and similar range in ($x$,$Q^2$) for proton and deuteron structure
function.
In Fig.1 we display the experimental data for proton (left panel)
and deuteron (right panel) structure functions chosen in our fit.
We have selected only NMC, E665 and SLAC sets of data \cite{F2_data}
for both proton and deuteron structure functions,
together amounting to 1833 experimental points:
901 for the proton structure function and
932 for the deuteron structure function.
According to the arguments presented above, we have
omitted BCDMS and HERA data in the fitting procedure but these will
be compared to our parametrization when discussing the quality of the
fit.

The deuteron structure function has been calculated as
\begin{equation}
F_2^d(x,Q^2) = {1 \over 2} [ F_2^p(x,Q^2) + F_2^n(x,Q^2) ] \; ,
\label{F2d}
\end{equation}
i.e. we have neglected all nuclear effects such as shadowing,
antishadowing due to excess mesons, Fermi motion, binding, etc,
which are known to be relatively small for the structure function
of the deuteron \cite{BK92,Z92,MT93}, one of the most
loosely bound nuclear systems.
In addition we have assumed isospin symmetry for the proton and
neutron quark distributions, i.e.
$u_n(x,Q^2) = d_p(x,Q^2)$, $d_n(x,Q^2) = u_p(x,Q^2)$ and
$s_n(x,Q^2) = s_p(x,Q^2)$. The charm contribution, which in the GRV
parametrization \cite{GRV95} is due to the photon-gluon fusion, is in
practice negligible in the region of $x$ and $Q^2$ taken in the fit,
and therefore is omitted throughout the present analysis.

The results of the fit are summarised in Table 1.
Because in general $Q_0^2$, $Q_1^2$ and $Q_2^2$ can be
different, there are 4 independent free parameters of the model.
In order to limit the number of parameters we have imposed extra
conditions as specified in Table 1. A series of seven fits
has been performed. In all cases considered the number
of free parameters has been reduced to two: the cut-off parameter of
the form factor and $Q_0^2$.
Statistical and systematic errors were added in quadrature when
calculating $\chi^2$. Only data with $Q^2 >$ 0.25 GeV$^2$ were
taken in the fit which is connected with the domain of applicability
of the GRV parametrization. The values of the parameters found
are given in each case in parentheses below the value of
$\chi^2$ per degree of freedom. In addition to the combined
fit, which includes both proton $F_2^p$ and deuteron $F_2^d$ structure
function data, we show the result of the fit separately for proton
and deuteron structure functions. As can be seen from the table
fairly similar values of parameters are found for the proton and
deuteron structure function and the $\chi^2$ per degree of freedom is
slightly worse in the latter case which can be due to the omission
of nuclear effects as mentioned above. The best fit (No 1 in the table)
is obtained with $Q_1^2 = Q_2^2 = 0$ (fits of similar quality can
be obtained with very small values of $Q_1^2 \sim$ 0.1 GeV$^2$
and $Q_2^2 \sim$ 0.1 GeV$^2$). While the value of $\chi^2$
does not practically depend on the type of the form factor
(exponential vs. Gaussian), much larger value of $Q_0^2$
is found for the Gaussian ($Q_0^2$ = 0.84 GeV$^2$) than for
the exponential ($Q_0^2$ = 0.52 GeV$^2$) parametrization.
The value of $Q_0^2$ found here is smaller than in the original
Bade{\l}ek-Kwieci\'nski model \cite{BK_model} but larger than that
found by H1 collaboration in the fit to low-$x$ data \cite{H1_lowQ2}.

Although the resulting $\chi^2$
is similar in both cases, the $F_2^n(x) / F_2^p(x)$ ratio for $x
\rightarrow 1$ prefers the Gaussian form factor.
While the vector meson contribution with the exponential form factor
survives up to relatively large $x$,
with the Gaussian form factor it is negligible at large $x$..

For comparison the GRV parametrization of quark distributions alone
yields:\\
$\chi^2/N_{dof}$ = 9.74 (21.48) (proton structure functions),\\
$\chi^2/N_{dof}$ = 13.73 (32.99) (deuteron structure functions),\\
$\chi^2/N_{dof}$ = 11.77 (27.33) (combined data),\\
where the first numbers include target mass corrections and
for illustration in parentheses their counterparts without target
mass corrections are given. Clearly the inclusion of
the target mass effects is essential and only such results will be
discussed in the course of the present paper.

The agreement of the CKMT parametrization is comparable to that
obtained in our model. For instance for parametrization (b) in Table 2
in the second paper \cite{CKMT}, which includes new HERA data:\\
$\chi^2/N_{dof}$ = 2.22 (1.00) (proton structure functions),\\
$\chi^2/N_{dof}$ = 3.54 (3.59) (deuteron structure functions),\\
$\chi^2/N_{dof}$ = 2.89 (2.33) (combined data),\\
where in the parentheses we present $\chi^2$ for $Q^2 <$ 4 GeV$^2$
i.e. in the region of applicability of the CKMT parametrization.
\footnote{The number of experimental points is reduced then to
354 and 373 for proton and deuteron structure functions, respectively}
We note much better description of the proton data in
comparison to the deuteron data.
The agreement of the Donnachie-Landshoff parametrization \cite{DL94}
with the proton structure function data is of similar quality.

In Fig.2 we present for completeness a map of $\chi^2$ for our best fit
as a function of model parameters $Q_0^2$ and $\lambda$.
A well defined minimum of $\chi^2$ for $\lambda_G \approx$ 0.5 GeV
and $Q_0^2 \approx$ 0.85 GeV$^2$ can be seen.
The experimental statistical uncertainty of the obtained parameters
$\lambda_G$ and $Q_0^2$ is less than 1 \%.

Some examples of the fit quality can be seen in Fig.3
(x-dependence for different values of $Q^2$ = 0.585, 1.1, 2.0, 3.5
GeV$^2$) and in Figs.4, 5 ($Q^2$-dependence for different values of
Bjorken $x$ = 0.00127, 0.0125, 0.05, 0.10, 0.18, 0.35, 0.55, 0.75).
Shown are experimental data \cite{F2_data} which differ from
the nominal $Q^2$ or Bjorken $x$ specified in Fig.3, 4, 5 by less
than $\pm$ 3 \%. An excellent fit is obtained for $Q^2 >$ 4 GeV$^2$
(not shown in Fig.3), although the VDM contribution stays large
up to 10 GeV$^2$.
In comparison to the GRV parametrization (dashed line)
our model describes much better the region of small $Q^2 <$ 3 GeV$^2$,
especially at intermediate Bjorken $x$: 0.05 $< x <$ 0.4.
The CKMT model (long-dashed line), shown according to the philosophy
in \cite{CKMT} for $Q^2 <$ 10 GeV$^2$ gives a better fit at very
small Bjorken $x$. However, one may expect here a few more effects
which will be discussed below.
\footnote{The use of the
next-to-leading order structure functions \cite{GRV95} in our model
would improve the description of low-$x$ data, discussion of which
is left for a separate analysis.}
It is however slightly worse as far as
isovector quantities are considered, as will be discussed later.
There seems to be a systematically small (up to about 5 \%) discrepancy
between our model and the data for $Q^2 <$ 2 GeV$^2$ and $x$ = 0.1 - 0.3.
This is caused by some higher-twist effects due to
the production of the $\pi N$ \cite{S72} and $\pi \Delta$ exclusive
channels and will be discussed elsewhere.
A fit of similar quality is obtained in our model for the proton (left
panels) and deuteron (right panels) structure functions.
Rather good agreement of our model with the BCDMS data can be observed
in Figs.4 and 5 in spite of the fact that the data were not used in
the fitting procedure.

At very small $x <$ 0.01 the description of the data becomes
worse. This is partially due to the use of the leading order
approximation. The fit to the fixed-target data prefers
$\bar x \approx x$ and $\bar Q^2 \approx Q^2$
(see Table 1 and the discussion therein).
On the other hand, the HERA data would prefer
$Q_1^2 \ne $ 0 and $Q_2^2 \ne $ 0. If we included
the HERA data in the fit the description of the fixed target data
would become worse. There are no fundamental reasons for
the parameters in both regions to be identical.
In addition at very small $x$ other effects of
isoscalar character,
not included here, such as heavy long-lived fluctuactions of the incoming
photon \cite{Shaw} and/or BFKL pomeron effects \cite{BFKL},
may become important.

For illustration a VDM contribution modified by a form factor
(\ref{time_formfactor}) is shown separately by the short-dashed line.
The partonic component can be obtained as a difference between the solid
and VDM line. It can be seen
from Figs.3-5 that the partonic component decreases towards small
$Q^2$. This decrease is faster than one could
directly infer from the failure of the GRV parametrization at low $Q^2$
because in our model a part of the strength resides in the
VDM contribution.
The modified VDM contribution is sizeable for small values
of Bjorken $x$ and not too large $Q^2$ and survives up to relatively
large $Q^2$. At $Q^2 >$ 3.5 GeV$^2$ the structure functions in our model
almost coincide with those in the GRV parametrization despite that the
VDM term is still not small.
For $Q^2 \rightarrow \infty$ only
the partonic contribution survives and
$F_2(x,Q^2) \rightarrow F_2^{part}(x,Q^2) \rightarrow F_2^{GRV}(x,Q^2)$.

The deviations from the partonic model can be also studied in
the language of higher-twist corrections. Then
the structure function can be written formally as
\begin{equation}
F_2^{p/n}(x,Q^2) = F_2^{p/n,LT}(x,Q^2)
\left[ 1 + \frac{ c_2^{p/n}(x) }{Q^2} + \frac{c_4^{p/n}(x)}{Q^4} + ...
\right]
\; .
\label{ht_expansion}
\end{equation}
However, in empirical analyses one usually includes only one term
in (\ref{ht_expansion})
\begin{equation}
F_2^{p/n}(x,Q^2) = F_2^{p/n,LT}(x,Q^2)
\left[ 1 + \frac{c^{p/n}(x) }{Q^2} \right] \; .
\label{ht_trunc}
\end{equation}
In our model for $Q^2 \sim M_V^2, Q_0^2$ there are an infinite number
of active terms in the expansion of the structure function
(\ref{ht_expansion}). Therefore the coefficient $c^{p/n}$ (the same is
true for the deuteron counterpart $c^d$) in (\ref{ht_trunc}) becomes
effectively $Q^2$-dependent $c^{p/n}(x) = c^{p/n}(x,Q^2)$.

As an example in Fig.6 we show $c^p$ and $c^d$ as a function of
Bjorken $x$ for three different values of $Q^2$ = 2, 4, 8 GeV$^2$
in the range relevant for the analysis in \cite{VM92}.
In order to correctly compare our results for $c^p$ and $c^d$ with the
results of the analysis in \cite{VM92} the structure function
$F_2^{p/n,LT}(x,Q^2)$ in Eq.(\ref{ht_trunc}) will include complete
target mass corrections calculated according to Ref.\cite{GP76}.
A fairly similar pattern is obtained for $c^p$ and $c^d$ especially
at small $x$.
The rise of $c^p$ or $c^d$ for $x \rightarrow$ 1 is caused by
our treatment of the target mass corrections and partially by the VDM
contribution which survives in our model up to relatively large x.
We obtain small negative $c^p$ and $c^d$ for $x <$ 0.3 in agreement
with \cite{VM92}. The smallnest of $c^p$ and $c^d$ in our model
for x $<$ 0.3 is due to the cancellation of the positive VDM contribution
and a negative contribution caused by the external damping factor
$\frac{Q^2}{Q^2+Q_0^2}$ of the partonic contribution in
Eq.(\ref{F2_part}).
The CKMT parametrization \cite{CKMT}
(shown only in its applicability range for $Q^2$ = 2, 4 GeV$^2$)
provides a very good explanation of $c^{p}$. It predicts, however,
somewhat larger negative $c^{d}$ for x $<$ 0.3. This will have
some unwanted consequences for $c^p - c^n$ discussed below.

In Fig.7 we show $c^p-c^n$ for $Q^2$ = 2, 4 GeV$^2$
together with empirical results from \cite{NMC_ht}. Rather strong
$Q^2$-dependence of $c^p-c^n$ is observed.
Our model correctly describes the trend of the experimental data.
For comparison we show also the result obtained by means of
the CKMT parametrization (long-dashed lines) of the
structure functions which somewhat fails to reproduce the details of
the empirical results from \cite{NMC_ht}, especially for small
 Bjorken $x$.
To our best knowledge no other model in the
literature is able to describe quantitatively this subtle higher-twist
effect.

Our model seems to provide a very good description
of some isovector quantities. As an example in Fig.8 we present
$F_2^p(x,Q^2) - F_2^n(x,Q^2)$ at $Q^2$ = 4 GeV$^2$ obtained
in our model (solid lines for different form factors), as well as the
results obtained with the GRV parametrization (dashed line) and
in the CKMT model (long-dashed line).\footnote{
No evolution of the CKMT quark distributions was done, but it is
negligible between 2 $\rightarrow$ 4 GeV$^2$
for the nonsinglet quantity.
}
The NMC data \cite{NMC} prefer rather our model.
As a consequence of the imperfect description of the deuteron data
the CKMT model fails to describe the difference $F_2^p(x) - F_2^n(x)$
for $x <$ 0.3. The success of our model is related to the violation of
the Gottfried Sum Rule and/or $\bar d - \bar u$ asymmetry
which is included in our model explicitly. In contrast to our model
in the CKMT model for $Q^2 >$ 2 GeV$^2$ the Gottfried Sum Rule
$S_G = {1 \over 3}$.

\section{Conclusions and discussion}

\mbox{}\indent
We have constructed a simple model incorporating nonperturbative
structure of the nucleon and photon. Our model is a generalization of
the well known and successful Bade{\l}ek-Kwieci\'nski model
\cite{BK_model}. While the original Bade{\l}ek-Kwieci\'nski model
is by construction limited to a small-$x$ region, our model is intended
to be valid in much broader range. The original VDM model assumes
implicitly a large coherence length for the photon-hadron fluctuation,
i.e. assumes that the hadronic fluctuation is formed far upstreem of
the target.
When the fluctuation length becomes small the VDM is expected to
break-down.
This effect has been modelled by introducing an extra form factor.
As a result we have succeeded in constructing a physically motivated
parametrization of both proton and deuteron structure functions.
In comparison to the pure partonic models with QCD evolution our model
leads to a much better agreement at low $Q^2$ in a broad range of $x$.

With only two free parameters we have managed to describe well the
transition from the high- to low-$Q^2$ region simultaneously for the
proton and deuteron structure functions. Our analysis of
the experimental data indicates that the QCD parton model
begins to fail already at $Q^2$ as high as about 3 GeV$^2$.
This value is larger than commonly believed.

In our discussion we have omitted the region of the HERA data.
In our opinion the physics there may be slightly more complicated.
Other effects of isoscalar character,
not included in our analysis,
for example the BFKL pomeron effects \cite{BFKL},
may become important.

In contrast to other models in the literature we obtain a very good
description of the NMC $F_2^p - F_2^n$ data \cite{NMC} where the
previously mentioned isoscalar effects cancel.

Recently an intriguing, although small, difference between
$\bar d - \bar u$ asymmetry obtained from recent E866 Drell-Yan data
\cite{E866} and muon DIS NMC data \cite{NMC} has been observed.
At least part of the difference can be understood in our model.
We expect for $Q^2$ smaller than about 4 GeV$^2$ an
extra $Q^2$ dependence of some parton model sum rules.
We predict a rather strong $Q^2$ effect for the integrand of the
Gottfried Sum Rule where in the first approximation the VDM contribution
cancels.

Recently in the literature there has been sizeable activity towards
a better understanding of higher-twist effects.
Some were estimated within the operator product expansion,
some in terms of the QCD sum rules. It is, however,
rather difficult to predict the absolute normalization of the
higher-twist effects.
Our model leads to relatively large higher-twist contributions.
For some observables, like structure functions, they almost cancel.
For other observables, like $F_2^p - F_2^n$, the cancellation is not
so effective. Our model provides specific higher-twist effects not
discussed to date in the literature. This will be a subject of a
future separate analysis.

\vskip 1cm

{\bf Acknowledgments:} \\
We are especially indebted to J. Kwieci\'nski for valuable
discussions and suggestions and J. Outhwaite for careful reading
of the manuscript. We would also like to thank C. Merino
for the discussion of the details of the CKMT model.
This work was supported partly by the German-Polish exchange program,
grant No. POL-81-97.

\newpage


\small{

\begin{table}

\caption{\it A compilation of the results obtained from our fit.
The $\chi^2$ per degree of freedom are given in first lines
whereas the pairs of numbers in second lines are the parameters
($\lambda$ (GeV), $Q_0^2$ (GeV$^2$))
found in the fit.}

\vskip 0.5cm

\begin{tabular}{|l|c|c|c|c|c|c|}
\hline
\hline
\multicolumn{1}{|c|}{}
& \multicolumn{3}{c|}{exponential}
& \multicolumn{3}{c|}{Gaussian} \\
\hline
case  & $F_2^p \  and \  F_2^d$ & $F_2^p$  & $F_2^d$ &
        $F_2^p \  and \  F_2^d$ & $F_2^p$  & $F_2^d$ \\
\hline
\hline
1) $Q_1^2 = Q_2^2 = 0$
 & 2.39        & 2.13        & 2.63        & 2.38        & 1.90        & 2.66        \\
 & (0.31,0.52) & (0.30,0.51) & (0.31,0.53) & (0.50,0.84) & (0.49,0.77) & (0.50,0.87) \\
\hline
\hline
2a) $Q_1^2 = Q_0^2$
 & 3.24        & 2.79        & 3.68        & 2.79        & 2.27        & 3.15        \\
 & (0.34,0.51) & (0.33,0.50) & (0.34,0.51) & (0.53,0.77) & (0.53,0.77) & (0.53,0.84) \\
\hline
2b) $Q_1^2 =$0.5 GeV$^2$
 & 3.20        & 2.77        & 3.61        & 2.53        & 2.05        & 2.85        \\
 & (0.34,0.53) & (0.34,0.52) & (0.34,0.53) & (0.52,0.82) & (0.52,0.78) & (0.52,0.85) \\
\hline
\hline
3a) $Q_2^2 = Q_0^2$
 & 4.13        & 3.46        & 4.74        & 3.60        & 3.14        & 4.02        \\
 & (0.37,0.43) & (0.36,0.44) & (0.37,0.43) & (0.56,0.66) & (0.57,0.66) & (0.56,0.66) \\
\hline
3b) $Q_2^2 =$0.5 GeV$^2$
 & 4.50        & 3.67        & 5.24        & 2.88        & 2.48        & 3.24        \\
 & (0.40,0.47) & (0.39,0.48) & (0.39,0.45) & (0.56,0.73) & (0.56,0.72) & (0.55,0.74) \\
\hline
\hline
4a) $Q_1^2 = Q_2^2 = Q_0^2$
 & 5.92        & 4.96        & 6.79        & 5.49        & 4.74        & 6.20        \\
 & (0.38,0.40) & (0.38,0.41) & (0.38,0.38) & (0.58,0.60) & (0.58,0.60) & (0.57,0.60) \\
\hline
4b) $Q_1^2 = Q_2^2$
 & 7.06        & 5.67        & 8.32        & 4.26        & 3.63        & 4.84        \\
 =0.5 GeV$^2$
 & (0.43,0.45) & (0.43,0.47) & (0.42,0.43) & (0.57,0.70) & (0.58,0.70) & (0.57,0.70) \\
\hline
\hline
\end{tabular}

\end{table}

}



\begin{figure}
\epsfig{file=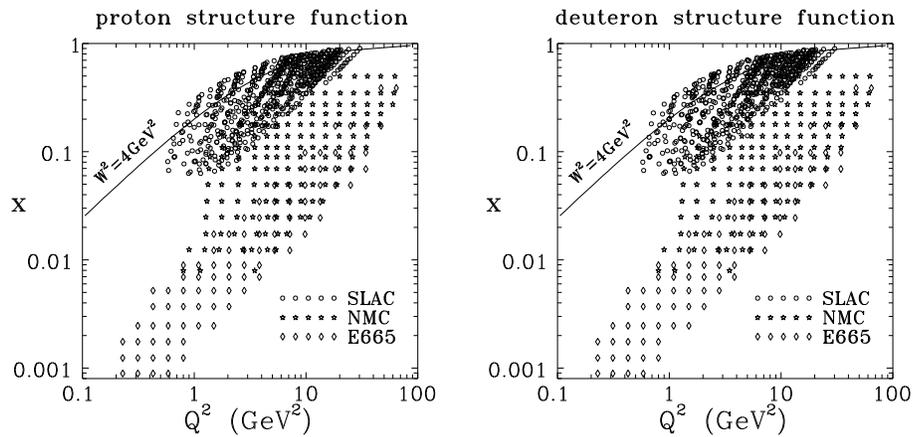,angle=0,height=100.mm,
bbllx=0, bblly=50, bburx=150, bbury=120}
\caption{\it
The presentation in the ($x,Q^2$) plane of the experimental data
included in the fit for proton (left panel) and deuteron (right panel)
structure functions.
For reference shown are lines with constant $W$ = 2 GeV which
conventionally separate resonance and DIS regions.
}
\label{data}
\end{figure}

\begin{figure}
\epsfig{file=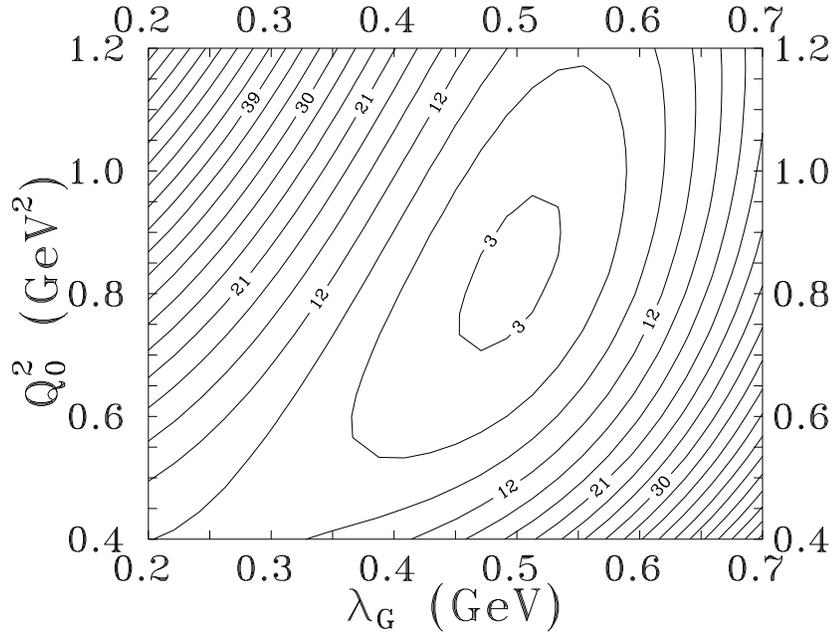,height=100mm,angle=0,
        bbllx=0,bblly=0.,bburx=100,bbury=120}
\caption{\it
A map of the $\chi^2$ per degree of freedom in the ($\lambda$, $Q_0^2$)
space of the model parameters for combined fit to
the proton and deuteron structure functions with
the Gaussian form factor.
}
\label{map}
\end{figure}


\begin{figure}
\begin{center}
\mbox{
\epsfysize 15.0cm
\epsfbox{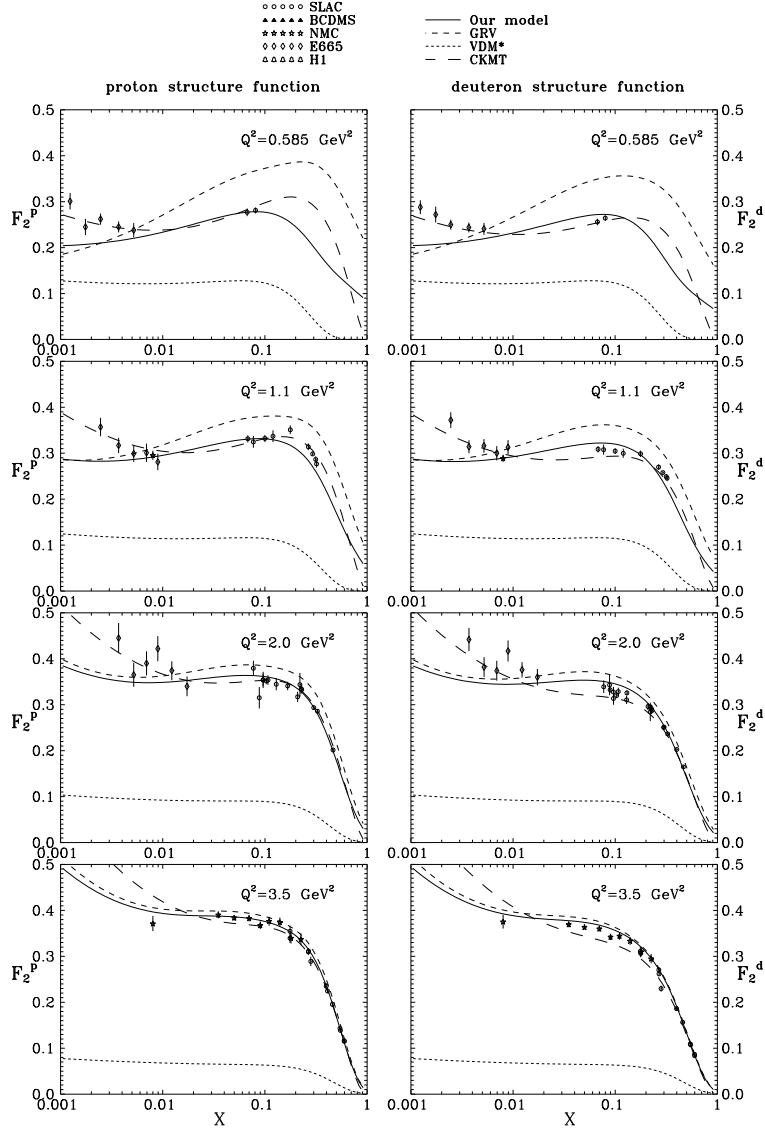}
}
\end{center}
\caption{\it
Comparison of the model results with experimental data for
$F_2^p$ (l.h.s.) and $F_2^d$ (r.h.s.)
as a function of Bjorken $x$ for different values of
$Q^2$ = 0.585, 1.1, 2.0, 3.5 GeV$^2$ .
The solid line corresponds to our full model with the Gaussian form
factor.
We present also the modified VDM contribution (short-dashed) and for comparison
also the result obtained with GRV parametrization \protect\cite{GRV95}
(corrected for target mass effects) of quark distributions (dashed line)
and that of the CKMT model \protect\cite{CKMT} (long-dashed line) .
}
\label{f2_x}
\end{figure}


\begin{figure}
\begin{center}
\mbox{
\epsfysize 15.0cm
\epsfbox{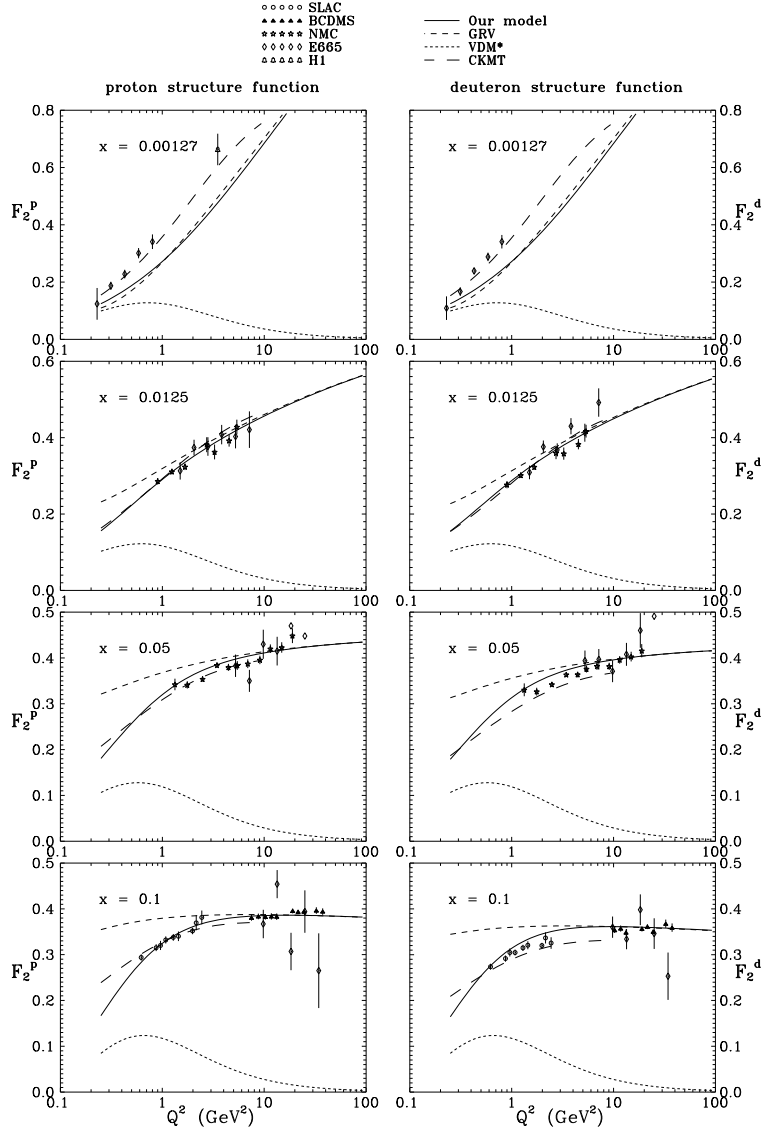}
}
\end{center}
\caption{\it
The same as in Fig.3 as a function of $Q^2$
for different values of
$x$ = 0.00127, 0.0125, 0.05, 0.10.
}
\label{f2_q2}
\end{figure}


\begin{figure}
\begin{center}
\mbox{
\epsfysize 15.0cm
\epsfbox{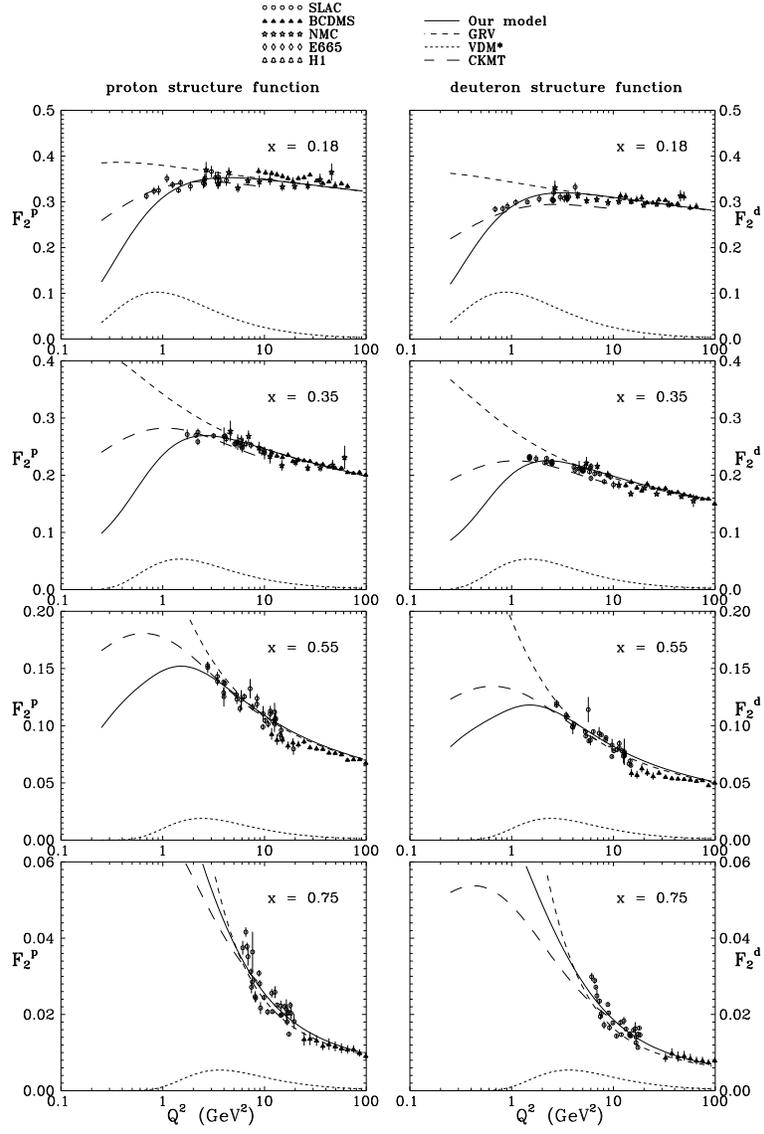}
}
\end{center}
\caption{\it
The same as in Fig.4 for different values of
$x$ = 0.18, 0.35, 0.55, 0.75.
}
\label{f2_q2}
\end{figure}


\begin{figure}
\epsfig{file=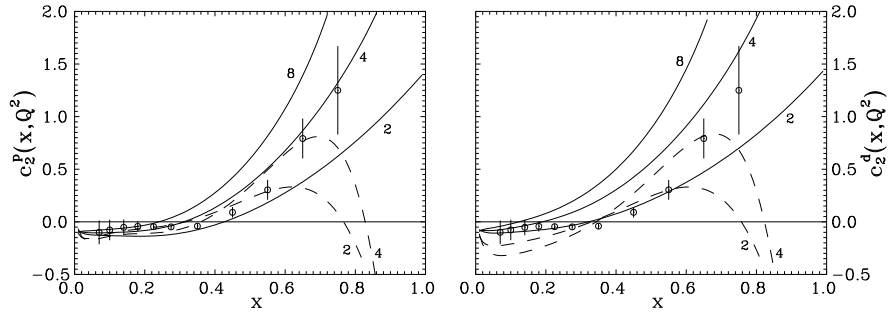,height=100mm,angle=0,
        bbllx=0,bblly=0,bburx=250,bbury=120}
\caption{\it
The twist-four coefficients $c^p$ and $c^d$ defined by Eq.
(\protect\ref{ht_trunc})
obtained in our model for $Q^2$ = 2,4,8 GeV$^2$. The coefficients
obtained from the CKMT parametrization are shown by the long-dashed
line for $Q^2$ = 2,4 GeV$^2$. The experimental data are taken
from \protect\cite{VM92}.
}
\label{c2pd}
\end{figure}


\begin{figure}
\epsfig{file=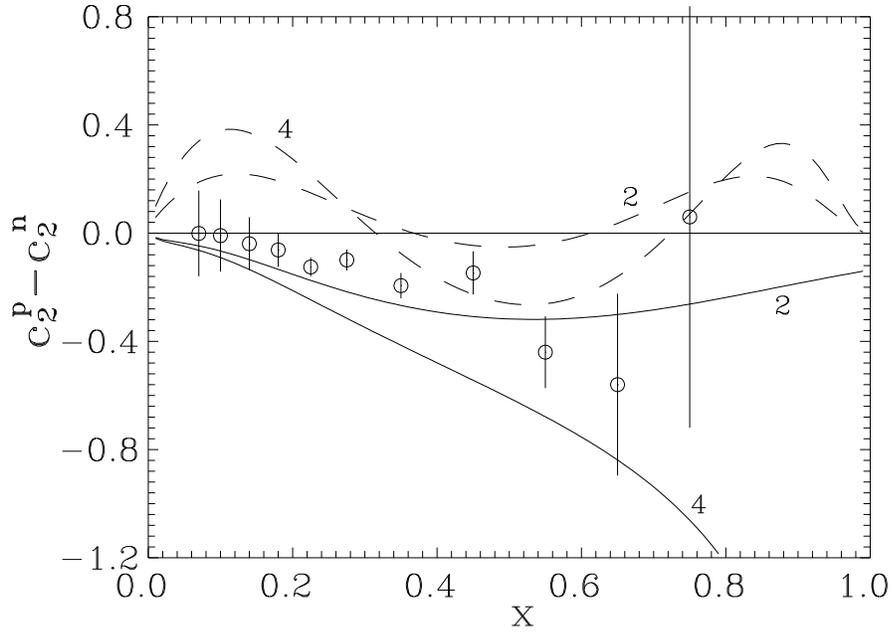,height=100mm,angle=0,
        bbllx=0,bblly=0,bburx=250,bbury=100}
\caption{\it
The difference of twist-four coefficients $c^p-c^n$. The data points are
from Ref. \protect\cite{NMC_ht}. The solid lines are the results of our
model while the dashed lines are obtained with the CKMT-parametrization.
In both cases the two curves are for $Q^2$ = 2,4 GeV$^2$.
}
\end{figure}


\begin{figure}
\epsfig{file=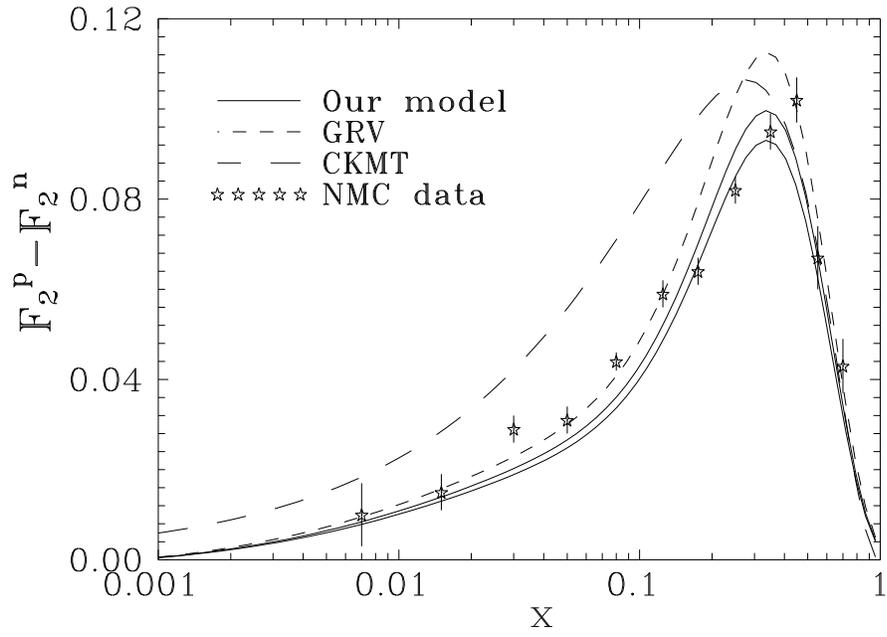,height=100mm,angle=0,
        bbllx=0,bblly=50,bburx=250,bbury=150}
\caption{\it
$F_2^p(x,Q^2) - F_2^n(x,Q^2)$ at $Q^2$ = 4 GeV$^2$ compared to the NMC
data. The upper solid line corresponds to our model with the exponential
form factor, the lower solid line to our model with the Gaussian
form factor, the dashed line to the GRV parametrization
and the long-dashed line to the CKMT parametrization.
}
\label{f2p_f2n}
\end{figure}


\end{document}